\documentclass[12pt]{article}

\pdfoutput=1
\usepackage{graphicx}
\usepackage{amssymb}
\usepackage{amsmath}
\usepackage{color}
\usepackage{cite}
\usepackage{subfigure}
\usepackage[colorlinks=true,linkcolor=blue,citecolor=blue]{hyperref}
\begin{document}

\newcommand{\beq}{\begin{equation}}
\newcommand{\eeq}{\end{equation}}

%\vskip 1cm

\begin{titlepage}

\begin{flushright}
TU-1103,~~IPMU20-0066, 
\end{flushright}

\vskip 2cm

\begin{center}

{\Large \bf
Primordial Black Holes from QCD Axion Bubbles
}

\vspace{1cm}

Naoya Kitajima$^{\,a,b}$
and
Fuminobu Takahashi$^{\,b,c}$ \\

\vskip 1.0cm

{\it
$^a$Frontier Research Institute for Interdisciplinary Sciences, Tohoku University, Sendai, 980-8578 Japan\\[2mm]
$^b$Department of Physics, Tohoku University, Sendai, 980-8578 Japan\\[2mm]
$^c$Kavli IPMU (WPI), UTIAS, The University of Tokyo, Kashiwa, Chiba 277-8583, Japan
}

\vskip 1.0cm

\begin{abstract}
We propose a scenario in which a strong Peccei-Quinn (PQ) symmetry breaking in the early universe results in large inhomogeneities  of the initial QCD axion field
value,  leading to the formation of very dense axion bubbles. Some of the axion bubbles subsequently collapse  into primordial black holes (PBHs). 
The spatially homogeneous part of the QCD axion explains dark matter of the universe, while the PBHs arising from the axion 
bubbles can explain the LIGO events or the seed of supermassive black holes. Interestingly, the mass of PBH is determined by the axion decay constant; 
for $f_a = 10^{17} (10^{16})$\, GeV, the PBH mass is heavier than about $10 (10^4) M_\odot$.
In addition, axion miniclusters are also formed from the axion bubbles more abundantly than PBHs, and their masses are expected to be heavier than in the usual scenario based on
the spontaneous breaking of the PQ symmetry after inflation.
\end{abstract}

\end{center}

\end{titlepage}

\newpage
%\tableofcontents

%\newpage
\vspace{1cm}

%%%%%%%%%%%%%%%%%%%%%%%%%%%%%%%%%%%%%%%%%%%%%%%%%%%%%%%%%%%%%%%%%%%%%%
\section{Introduction} \label{sec:intro}
%%%%%%%%%%%%%%%%%%%%%%%%%%%%%%%%%%%%%%%%%%%%%%%%%%%%%%%%%%%%%%%%%%%%%%
Dark matter in the universe is one of the greatest mysteries of particle physics and cosmology. If it is made up of unknown particles, 
then a new physics beyond the standard model (SM) is needed. For example, the 
Peccei-Quinn mechanism  \cite{Peccei:1977hh,Peccei:1977ur}, known as the solution to the strong CP problem, predicts a pseudo scalar called 
the QCD axion~\cite{Weinberg:1977ma,Wilczek:1977pj}, and it is one of the plausible candidates for dark matter.
Alternatively, all or part of the dark matter might be a primordial black hole (PBH)\cite{1967SvA....10..602Z,Hawking:1971ei,Carr:1974nx}, which 
forms from the direct collapse of the primordial cosmic fluid.

Various observations have shown that PBH cannot be the dominant component of dark matter unless its mass is within 
$M_{\rm PBH} \sim 10^{17}$--$10^{20}$ g (see \cite{Carr:2009jm,Carr:2020gox} for a review). However, even when PBH accounts for a small fraction
of dark matter, it can still leave interesting cosmological and astrophysical signatures; PBHs may form binaries and cause mergers that can be detected by gravitational wave observations. The estimated black hole mass is of ${\cal O}(10) M_\odot$ based on the black hole coalescence phenomenon detected
 by LIGO and Virgo~\cite{Abbott:2016blz,Abbott:2016nmj,Abbott:2017vtc,Abbott:2017oio}, which motivates the study of PBHs with such a mass\cite{Bird:2016dcv,Clesse:2016vqa,Sasaki:2016jop}.

Another problem is the origin of the supermassive black holes (SMBH) with mass ${\cal O}(10^9) M_\odot$ observed at redshift $z \sim 6-7$ \cite{Wu_2015}. 
This is because it is non-trivial for stellar black holes to grow into an SMBH at such a high redshift. 
On the other hand, PBHs form much earlier than the star-forming epoch and have enough time to grow into SMBHs, 
so the PBHs with mass $10^4-10^5 M_\odot$ may become seeds of SMBHs~\cite{Bean:2002kx}. 
It was also pointed out that PBH with a mass similar to that of SMBH could be produced directly from the beginning~\cite{Nakama:2016kfq}.

A number of cosmological scenarios have been proposed for the formation of PBHs. Many of such scenarios are
 based on a mechanism that increases density perturbation due to the inflaton dynamics (see Ref. \cite{Carr:2016drx} for a review). Then, the PBH is formed when
the enhanced density perturbations enter the Hubble horizon well after inflation. In this class of cosmological scenarios,
 it depends on the details of the inflaton dynamics on what scale and to what extent the density perturbations are enhanced, and therefore,
  the mass of PBH is usually a free parameter.

In this paper, we propose a scenario in which the PBH is formed by large inhomogeneities of the QCD axion with the PQ symmetry broken during inflation. In particular, we are interested in PBH masses of order or greater than  $10\, M_\odot$ in terms of explaining the LIGO events or the seeds of SMBHs. 
We consider a situation in which the PQ symmetry is strongly broken in the early universe, and the probability distribution of 
the initial misalignment angle is split to the two values close to $0$ and $\pi$. Such large deviation from the Gaussian distribution 
contrasts with the usual case where the density perturbation is approximated by a Gaussian distribution at the leading order. 
We assume that the probability distribution around $\pi$ is so suppressed that its volume fraction is very small and rare, while 
 the initial angle is close to the origin for most of the space. 
When the cosmic temperature drops down to the QCD scale, the QCD axion starts to oscillate near the origin for most of the space, which explains the dark matter.
On the other hand, although it is very rare, the QCD axion near $\pi$ acquires a very large energy density after the start of the oscillation. 
 We call this high density region {\it an axion bubble}. As we shall see, some of the axion bubbles collapse into PBHs or form miniclusters. Interestingly, our scenario predicts a PBH of mass $\gtrsim {\cal O}(10) M_\odot$, which is determined by the QCD scale and the axion decay constant. Thus, our scenario naturally leads to the mass scale of PBH needed to explain  the LIGO/Virgo event and the seeds of SMBHs, while at the same time explaining dark matter by the QCD axion.

Lastly let us mention related works in the literature. While we are considering a scenario in which PBH is generated from large inhomogeneities of dark matter,  PBH can also be generated from the baryon asymmetry.
The PBH formation in the Affleck-Dine mechanism was studied in Refs.~\cite{Dolgov:2008wu,Blinnikov:2016bxu,Hasegawa:2017jtk,Hasegawa:2018yuy,Kawasaki:2019iis}. In this case, 
large inhomogeneity of the Affleck-Dine field is realized by the inflationary fluctuation and nontrivial post-inflationary dynamics in the presence of
the multiple vacua. Then, it results in highly inhomogeneous distribution of the baryon asymmetry, leading to isolated high baryon bubbles. 
After the QCD phase transition, the baryon density becomes significant in the high baryon bubbles, some of which
collapse into PBHs if a certain criterion is met.
Since the PBH formation occurs after the QCD transition, the PBH mass is heavier than ${\cal O}(M_\odot)$, and thus it can potentially explain the LIGO/Virgo events~\cite{Blinnikov:2016bxu,Hasegawa:2017jtk,Hasegawa:2018yuy,Dolghov:2020hjk}
or seeds of SMBHs~\cite{Kawasaki:2019iis}.
 In addition, the PBH formation from the QCD axion with an alternative setup was studied in \cite{Ferrer:2018uiu}, in which the PQ symmetry is broken after inflation. This scenario focuses on the axion as a subdominant part of the dark matter with mass about meV and PBHs are seeded by the string-wall system. Thus, our scenario is complementary to this scenario.

The rest of this paper is organized as follows. In Sec.~\ref{sec:QCDaxion}, we review the conventional scenario of the QCD axion and its modification 
due to the Witten effect as an example of the early-time PQ breaking.
In Sec.~\ref{sec:PBH}, we show the formation of the axon bubbles, PBHs, and miniclusters in our scenario. 
Discussion and conclusions are given in Sec.~\ref{sec:conc}.

%%%%%%%%%%%%%%%%%%%%%%%%%%%%%%%%%%%%%%%%%%%%%%%%%%%%%%%%%%%%%%%%%%%%%%
\section{The QCD axion with the early-time PQ breaking} \label{sec:QCDaxion}
%%%%%%%%%%%%%%%%%%%%%%%%%%%%%%%%%%%%%%%%%%%%%%%%%%%%%%%%%%%%%%%%%%%%%%

\subsection{Vacuum realignment mechanism}

First, let us review the axion abundance in the conventional scenario.
When the cosmic temperature is much higher than the QCD scale, the axion remains almost massless.
As the temperature approaches to the QCD scale,
the axion acquires a mass that depends on the temperature \cite{Borsanyi:2016ksw},
\begin{equation}
\label{maT}
m_a(T) \simeq m_a \left(\frac{T_*}{T}\right)^{4.08},
\end{equation}
for $T>T_* = 150$ MeV, which asymptotes to $m_a \simeq 5.7\, \mu{\rm eV} \,(10^{12}\,{\rm GeV}/f_a)$ in the low temperature. 
Here $f_a$ denotes the decay constant of the QCD axion. 
When the axion mass becomes comparable to the Hubble parameter, the axion starts to oscillate. 
For $f_a \lesssim 10^{17}$\,GeV, this occurs when the axion mass is well approximated by (\ref{maT}).
We define the temperature $T_1$ by $3H(T_1) = m_a(T_1)$, and it is given by
\begin{equation}
T_1 \simeq 0.96\,{\rm GeV} \left(\frac{g_{*1}}{60}\right)^{-0.082} \left(\frac{f_a}{10^{12}\,{\rm GeV}}\right)^{-0.16},
\end{equation}
where $g_{*1}$ denotes the effective relativistic degrees of freedom at $T=T_1$.
Then, one obtains the final axion abundance via $\rho_a/s = \kappa m_a(n_a/s)_{T=T_1}$,
where $\kappa$ is the numerical fudge factor and we set $\kappa = 1.5$. 
In other words, the current density parameter of the axion is
\begin{equation}
\Omega_a h^2 \simeq 0.15\left(\frac{g_{*1}}{60} \right)^{-0.42} \left(\frac{f_a}{10^{12}\,{\rm GeV}} \right)^{1.2} \theta_i^2 F(\theta_i),
\end{equation}
where $\theta_i$ is the initial misalignment angle and  $F(\theta_i) = [\ln(e/(1-\theta_i^2/\pi^2))]^{1.2}$ is the anharmonic correction factor \cite{Visinelli:2009zm}.
Thus, one can explain the observed DM abundance for e.g. $f_a \sim 10^{12}$\,GeV for $\theta_i \sim 1$, and $f_a \sim 10^{16}$\,GeV for $\theta_i \sim
4 \times 10^{-3}$. Recently it was pointed out that the required small values of $\theta_i$ for large $f_a$ 
can be explained if the inflation scale is below the QCD scale~\cite{Graham:2018jyp,Guth:2018hsa}. In the following, however, we will not
worry about the fine-tuning of $O(10^{-3})$ required to explain the dark matter abundance by the QCD axion.

\subsection{Impact of the early-time PQ breaking on the QCD axion dynamics}

\subsubsection{Overview}

Here and in what follows, we assume that the PQ symmetry is broken before/during inflation. In this case, the axion abundance depends on the initial angle $\theta_i$ and the decay constant $f_a$.
The initial angle $\theta_i$ is often treated as a free parameter, since the QCD axion is massless during
inflation and any values of $\theta_i$ are considered to be equally 
probable. Note that this scenario unavoidably predicts the isocurvature perturbation which is severely constrained by the observation. Although one can simply adopt the low-scale inflation, post-inflationary classical dynamics can also suppress the isocurvature perturbation. For instance, an additional PQ breaking term can dynamically suppress the isocurvature perturbation. In fact, since the PQ symmetry is only an approximate global symmetry, it is not surprising that it was largely
broken in the early universe. In that case, the breaking must be small enough in the present universe 
for the PQ mechanism to be the solution to the strong CP problem.
A number of such scenarios have been proposed, such as a stronger QCD in the early universe due to 
the Higgs' large expectation value~\cite{Dvali:1995ce,Choi:1996fs,Banks:1996ea,Jeong:2013xta}, a larger scale of the 
spontaneous PQ symmetry breaking that resulted 
in a larger explicit breaking expressed by a higher dimensional term~\cite{Chiba:2003vp,Takahashi:2003db,Higaki:2014ooa,Co:2019jts}
 a hidden non-Abelian gauge symmetry 
that confines in the early universe~\cite{Takahashi:2015waa},
and the Witten effect of monopoles in hidden sectors~\cite{Kawasaki:2015lpf,Nomura:2015xil,Kawasaki:2017xwt}.
It should be emphasized that such scenarios can result in not only the suppression of the isocurvature fluctuation but also the tuning of the initial angle and thus predict relatively large $f_a $ ($\sim 10^{16}$ GeV), which is compatible with the string theory. Furthermore, 
such extra PQ breaking term generically has multiple vacua.
Then, the axion may fall into different vacua depending on the initial position, as we shall see  in Sec.~\ref{sec:PBH}.

In the following, we consider a scenario in which the PQ symmetry is temporarily broken by a large amount in the early universe. In this scenario, 
the axion is massless during inflation, and therefore it acquires vacuum quantum fluctuations during inflation. Then, after the end of inflation, an extra PQ symmetry-breaking effect  kicks in, which generates a potential for the axion. When the effective mass of the axion is greater than the Hubble parameter, the axion rolls down to the nearest minimum. This potential is assumed to be transient and to disappear before the QCD phase transition. Therefore, the final abundance of the axon is determined by the usual misalignment mechanism. Note that the initial misalignment angle $\theta_i$ is dynamically set by the transient PQ breaking potential. This also implies that the axion dynamics suppresses the isocurvature perturbation. In the rest of this section, we consider the Witten effect as an explicit example of the early transient PQ symmetry breaking.

\subsubsection{Witten effect}

Here we consider the Witten effect~\cite{Witten:1979ey},  which can suppress both the axion abundance and isocurvature.
To this end we couple the QCD axion $\phi$ to a hidden U(1)$_{\rm H}$ gauge boson as
\beq
{\cal L}_{H} = - \frac{\alpha_H}{16 \pi} \left(N_H \frac{\phi}{f_a}  + \theta_H \right) \epsilon_{\mu\nu\rho\sigma}
F_H^{\mu \nu} F_H^{\rho \sigma},
\label{eq:theta}
\eeq
where $\alpha_H$ is the hidden fine-structure constant, $\epsilon_{\mu \nu \rho \sigma}$ is the totally antisymmetric tensor,
$F_{H}^{\mu \nu} = \partial^\mu A_{H}^{\nu} - \partial^\nu A_{H}^{\mu} $ is the field strength of the hidden gauge boson 
$A_{H}^{ \mu}$,
 $N_H$ is an integer called the domain wall number, and $\theta_H$ is the $\theta$-parameter of the U(1)$_H$ gauge
 symmetry. Note that we choose the origin of the axion so that the CP phase of the strong interaction disappears there.
 Thus, $\theta_H$ is generically nonzero.
 
Let us first focus on the $\theta$-term without the axion.
It is just a surface term and is usually discarded because it does not contribute to the equation of motion. However, when a monopole is present, $\theta_H$ has a physical meaning. That is, a monopole acquires an electric charge proportional to $\theta_H$, and therefore it becomes a dyon. 
This is the Witten effect~\cite{Witten:1979ey}. Now, in the presence of the axion,
the $\theta$-parameter becomes dynamical, and the monopole acquires an electric charge proportional to the axion displacement from the CP-conserving point.  Since the dyon is heavier than the monopole, the interaction (\ref{eq:theta}) generates
a potential of the axion. If monopoles and anti-monopoles are uniformly distributed on average in the universe, then
the axion acquires a mass~\cite{Fischler:1983sc}.
\beq
m_{a, W}^2(t) = \frac{\alpha_H N_H}{16 \pi^2} \frac{n_M(t)}{r_c f_a^2},
\eeq
where $n_M$ is the number density of monopoles and anti-monopoles, and $r_c$ is the radius of the monopole core. 
In the case  of the 't Hooft-Polyakov monopole~\cite{tHooft:1974kcl,Polyakov:1974ek}, the core radius is given by $r_c^{-1} \simeq \alpha_H M_M$, where
$M_M$ is the monopole mass. 
The minimum of the mass term is located at $\phi = \phi_{\rm min}^{(n)}$ given by
\beq
\phi_{\rm min}^{(n)} = ( - \theta_H + 2 \pi n) \frac{f_a}{N_H},
\label{minW}
\eeq
where $n$ is an integer (see Fig.~\ref{fig:pot}). 
Since the axion mass squared due to the Witten effect, $m_{a, W}^2(t)$,  is proportional to the monopole number density, it decreases as 
\beq
m_{a,W}^2(t) \propto n_M(t) \propto a(t)^{-3},
\label{mawa}
\eeq
where $a(t)$ is the scale factor. Here we assume that the number of the monopoles
in the comoving volume is conserved after the formation. 
 In the radiation dominated era, the axion mass decreases more slowly than the
Hubble parameter, while in the matter dominated era, both decrease at the same rate.
Then, the axion starts to oscillate about one of the potential minima when the mass becomes comparable to the Hubble parameter in 
radiation domination.

If the U(1)$_H$ symmetry is subsequently spontaneously broken, there appear cosmic strings whose ends are attached to monopoles and anti-monopoles. Due to the tension of the strings, monopoles and anti-monopoles
are attracted to each other, dissipating its kinetic energy into the ambient plasma, until they collide and annihilate with each other.
The produced hidden Higgs bosons decay into the standard model particles
through e.g. a Higgs portal coupling. After the monopole-anti-monopole annihilation, the extra axion mass disappears, and
the axion remains massless until the temperature drops down to the QCD scale. 
If the duration of the axion oscillation due to the Witten effect lasts sufficiently long, the axion oscillation amplitude can be suppressed significantly,
and the initial axion field value is set to be around one of the potential minima given by Eq.~(\ref{minW}). The axion isocurvature perturbation is 
also  suppressed in this process.

In the following, we show that the axion can be indeed stabilized in the very vicinity 
of one of the potential minima by  the Witten effect.
We basically follow the argument of Ref.~\cite{Nomura:2015xil}, but we will take account of an important factor which was
missed in their analysis. Without loss of generality, we consider the axion oscillating around the 
minimum with $n=0$. To be explicit, we assume high-scale inflation, $H_{\rm inf} \sim 10^{13}$\,GeV,
high reheating temperature, $T_{\rm RH} \sim 10^{15}$\,GeV, and the monopole production right after
the reheating. Note that such a high scale inflation, which is usually in a strong tension with the axion 
model due to too large isocurvature fluctuations, can be made consistent thanks to the Witten effect. 

The axion starts to oscillate when $m_{a,W}$ becomes comparable to the Hubble parameter.
The plasma temperature, $T = T_{\rm osc}$, at the commencement of oscillation is
\beq
T_{\rm osc} \simeq 1.5 \times 10^3 N_H \alpha_H^2 \left(\frac{\rho_M}{s}\right) \left(\frac{10^{16}{\rm\,GeV}}{f_a}\right)^2,
\eeq
where $\rho_M$ is the energy density of monopoles and anti-monopoles, $s$ is the entropy density before the monopoles
dominate,
and we assume radiation domination. For the parameters of our interest, monopoles come to dominate the universe
before annihilation. The temperature at which the monopole becomes the dominant component of the universe is defined 
as follows.
\beq
T_{\rm dom} \equiv \frac{4}{3} \frac{\rho_M}{s}.
\eeq
The monopole abundance depends on details of the phase transition, and its abundance is bounded as~\cite{Vilenkin:2000jqa}
\beq
10^4 {\rm\, GeV}   \left(\frac{M_M}{10^{15 }{\rm \,GeV}}\right) 
\left(\frac{T_c}{10^{15 }{\rm \,GeV}}\right)^3 \lesssim 
\frac{\rho_M}{s} \lesssim 10^8  {\rm\, GeV} \left(\frac{M_M}{10^{15 }{\rm \,GeV}}\right)^2,
\eeq
where the lower bound is the causality bound, the upper bound is determined by the annihilation process between monopoles
and antimonopoles,  $T_c$ is the temperature at the bubble nucleation, and we have used $g_* \sim 100$ and $\alpha_H \sim 4 \pi$
for a conservative estimate.  Thus, for the reference values of the parameters, we have
$T_{\rm dom} \sim 10^{4-8}$\,GeV and $T_{\rm dom}/T_{\rm osc} \sim 10^{-3}$. Note that this is an order-of-magnitude estimate,
and the precise values of $T_{\rm dom}$ can vary depending on the choice of the parameters. 
When the monopoles become dominant, the universe evolves like a matter-dominated universe, and when the monopoles 
annihilate and decay, it becomes radiation dominant. Let us express the decay temperature by $T_{\rm dec}$. 
Here and in what follows, the subscripts, ``osc", ``dom", and ``dec" mean that the variable is evaluated when
the temperature of the radiation dominated universe 
is equal to $T= T_{\rm osc}, T_{\rm dom},$ and $T_{\rm dec}$, respectively.

The evolution of the axion can be divided into two  periods. The first is when the oscillation is caused by the mass term due to the Witten effect, and the second is when the monopoles annihilate and the axion mass is reduced. Only the former was considered in Ref.~\cite{Nomura:2015xil}, and the effect of the latter was neglected. However, as we will see below, the latter is important because 
it has the effect of increasing the oscillation amplitude of the axion. During the first period, the axion number density in the comoving volume is conserved, 
\beq
m_{a,W}(t) \Phi^2 \propto a(t)^{-3}
\eeq
where $\Phi$ denote the oscillation amplitude, and $a$ is the scale factor. At the commencement of  the oscillations,
we have $\Phi_{\rm osc} = \phi_{\rm osc} - \phi_{\rm min}^{(0)}$.
On the  other hand, the axion mass due to the Witten effect scales as Eq.~(\ref{mawa}).
Thus, $\Phi(t) \propto a(t)^{-3/4}$. 

\begin{figure}[tp]
\centering
\subfigure[]{
\includegraphics [width = 7.5cm, clip]{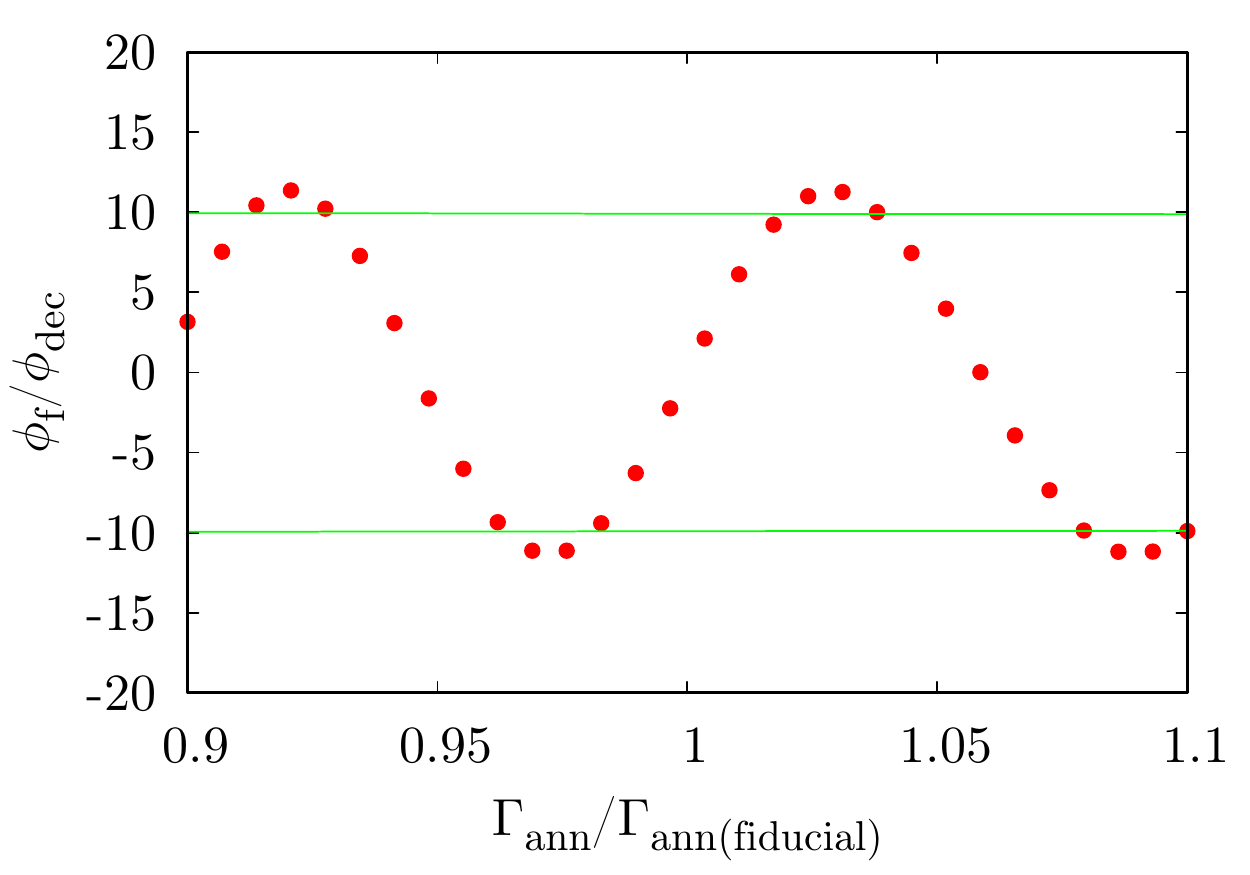}
\label{subfig:amplitude}
}
\subfigure[]{
\includegraphics [width = 7.5cm, clip]{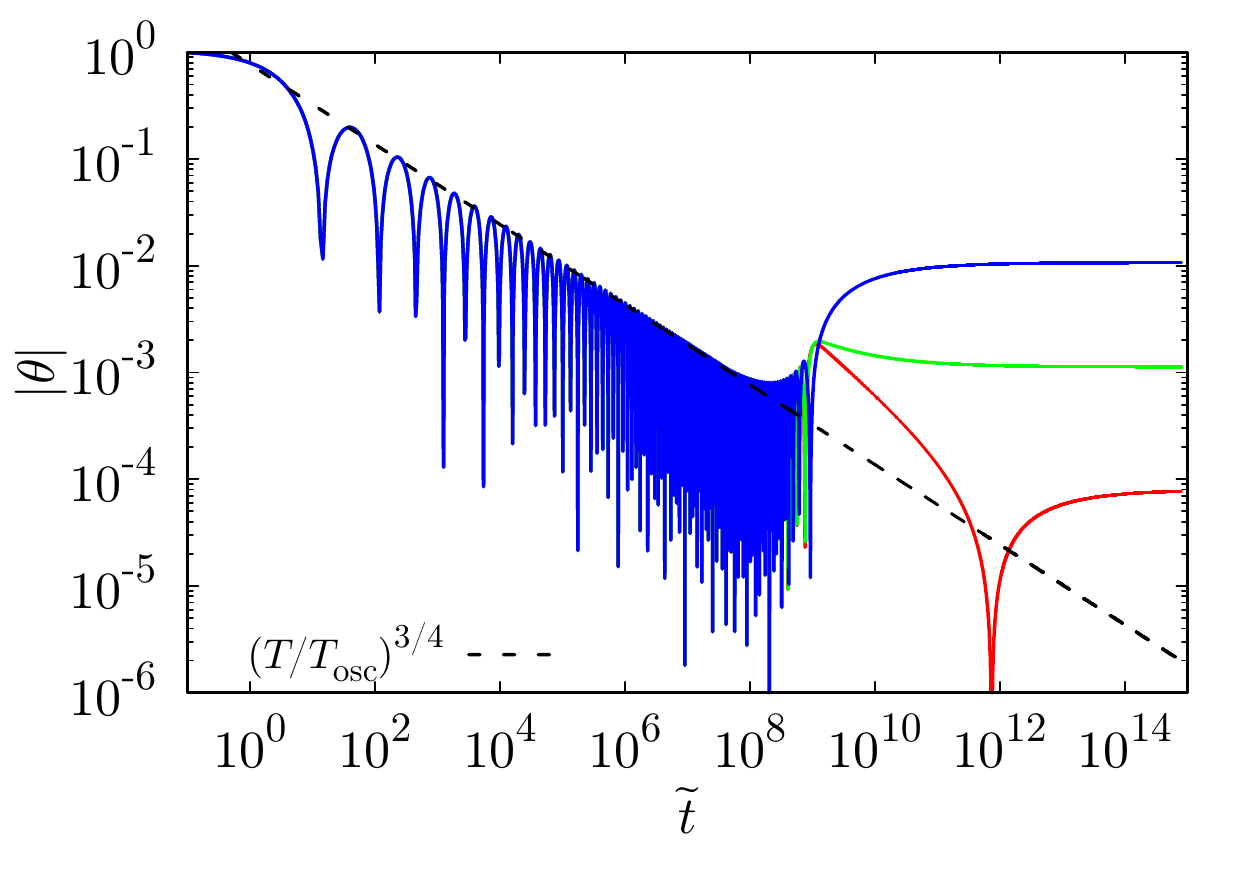}
\label{subfig:evolve}
}
\caption{
{\it Left}: the final amplitude of the axion field value ($\phi_f$) divided by the one at the decay 
as a function of the annihilation rate of the monopole, $\Gamma_{\rm ann}$ (normalized by some fiducial value). 
Here we consider the oscillation around $\phi_{\rm min}^{(0)}$ and set it to the origin for simplicity. 
The green horizontal lines show the maximal enhancement factor, $\pm(m_{a,W}(t_{\rm dec})/H_{\rm dec})^{1/2}$. {\it Right}: the evolution of the axion field value ($\theta=\phi/f_a$) for $\Gamma_{\rm ann}/\Gamma_{\rm ann(fiducial)}=1$ (red), 1.002 (green), 1.025 (blue). 
The horizontal axis is the dimensionless time variable, $\tilde{t} = m_{a,W}(t_{\rm osc})\,t$.
}
\label{fig:evolve}
\end{figure}

During the second period, the axion mass decreases and the oscillation amplitude increases. 
The axion stops to oscillate when the axion mass becomes comparable to the Hubble parameter, and then,
the axion field freezes. If the annihilation process takes of order the Hubble time,  the oscillation amplitude is enhanced maximally
by a factor of $(m_{a,W}(t_{\rm dec})/H_{\rm dec})^{1/2} = \left(a_{\rm dom}/a_{\rm osc}\right)^{1/4} $, since the axion number density in the 
comoving volume is conserved during this process. 
Here $m_{a,W}(t_{\rm dec})$ is the mass evaluated just before the
monopoles start to annihilate. We have numerically checked this effect as shown in Fig.~\ref{fig:evolve}.
The left panel shows the ratio of the final axion field value to the oscillation amplitude of the axion field
at the beginning of the annihilation. The horizontal green line shows the above enhancement factor.
Since the final field value depends on the oscillation phase when the annihilation starts, we varied the annihilation rate
(the horizontal axis). One can see that the numerical result is consistent with the enhancement factor, and our estimate
therefore provides an upper bound on the displacement from the potential minimum. The right panel shows the
time evolution of the axion field value $(\theta = \phi/f_a)$ as a function of time for representative choices of the
annihilation rate.

%%%%%%%%%%%%%%% MULTI-FIGURE  %%%%%%%%%%%%%%%
%%%%%%%%%%%%%%%%%%%%%%%%%%%%%%%%%%%%%%%

Combining the above two effects, we obtain the typical displacement of the axion from the potential minimum after the monopole annihilation
 as
\begin{align}
\Delta \theta 
&=\left(\frac{a_{\rm dom}}{a_{\rm osc}}\right)^{\frac{1}{4}} \left(\frac{a_{\rm dec}}{a_{\rm osc}}\right)^{-\frac{3}{4}} \frac{\Phi_{\rm osc}}{f_a},\\
&=\left(\frac{T_{\rm dec}}{T_{\rm dom}}\right)^{3/4} \left(\frac{T_{\rm dom}}{T_{\rm osc}}\right)^{\frac{1}{2}} \frac{\Phi_{\rm osc}}{f_a},
\end{align}
where we have assumed that the effective relativistic degrees of freedom is constant over the relevant period in the last
equality. The axion initial amplitude is therefore set as
\beq
\theta_i \simeq \theta_{\rm min}^{(0)} \pm \Delta \theta,
\eeq
where we defined $\theta_{\rm min}^{(n)} = \phi_{\rm min}^{(n)} /f_a$.

The axion can explain the right abundance of dark matter for $$\theta_i \simeq 4.2\times 10^{-3} \left(\frac{g_{*1}}{60} \right)^{0.21} \left(\frac{10^{16}\,{\rm GeV}}{f_a}\right)^{0.58}.$$
Thus, for $T_{\rm dec} \lesssim 10^{-3}\, T_{\rm dom}$, $\Phi_{\rm osc}/f_a \sim 1$, and $f_a \sim 10^{16}$\,GeV,
 the axion is stabilized close enough to the potential minimum. In addition, isocurvature fluctuations are sufficiently suppressed 
 because the fluctuation of the axion field evolves in the same way as the axion field in the quadratic potential.

%%%%%%%%%%%%%%%%%%%%%%%%%%%%%%%%%%%%%%%%%%%%%%%%%%%%%%%%%%%%%%%%%%%%%%
\section{PBH formation from the QCD axion} \label{sec:PBH}
%%%%%%%%%%%%%%%%%%%%%%%%%%%%%%%%%%%%%%%%%%%%%%%%%%%%%%%%%%%%%%%%%%%%%%

\subsection{Axion bubbles}
Here we show that the modification of the axion dynamics due to the early-time PQ symmetry breaking can lead to formation 
of the isolated high axion density regions. 
We consider the Witten effect explained in the previous section as an example of such extra PQ symmetry breaking.

\begin{figure}[tp]
\centering
\includegraphics [width=9cm,clip]{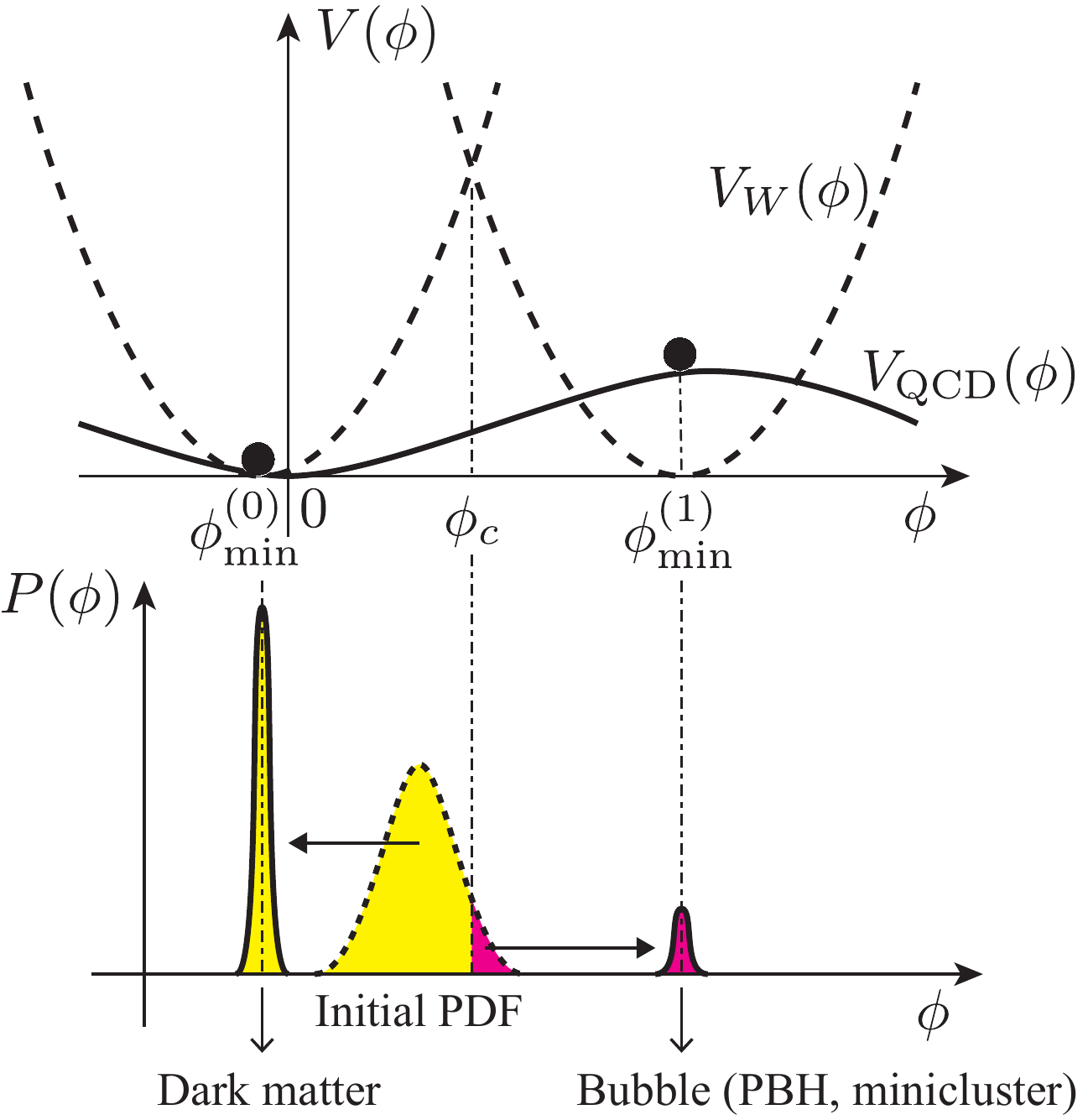}
\caption{
Schematic illustration of the axion potentials and the probability distribution. The solid and dashed lines represent 
 the potential from the QCD instanton effect ($V_{\rm QCD}$) and the Witten effect ($V_W$), respectively. 
}
\label{fig:pot}
\end{figure}

The axion potential due to the Witten effect is approximately given by $V_W(\phi)= m^2_{a,W}(\phi-\phi_{\rm min}^{(n)})^2/2$, which is shown by the dashed line in
 Fig.~\ref{fig:pot}. Here and in what follows we take $N_H=2$.
The Witten effect is assumed to disappear before the QCD phase transition when the conventional axion 
potential $V_{\rm QCD}(\phi)$ arises from non-perturbative QCD effects.

\begin{figure}[tp]
\centering
\includegraphics [width=9cm,clip]{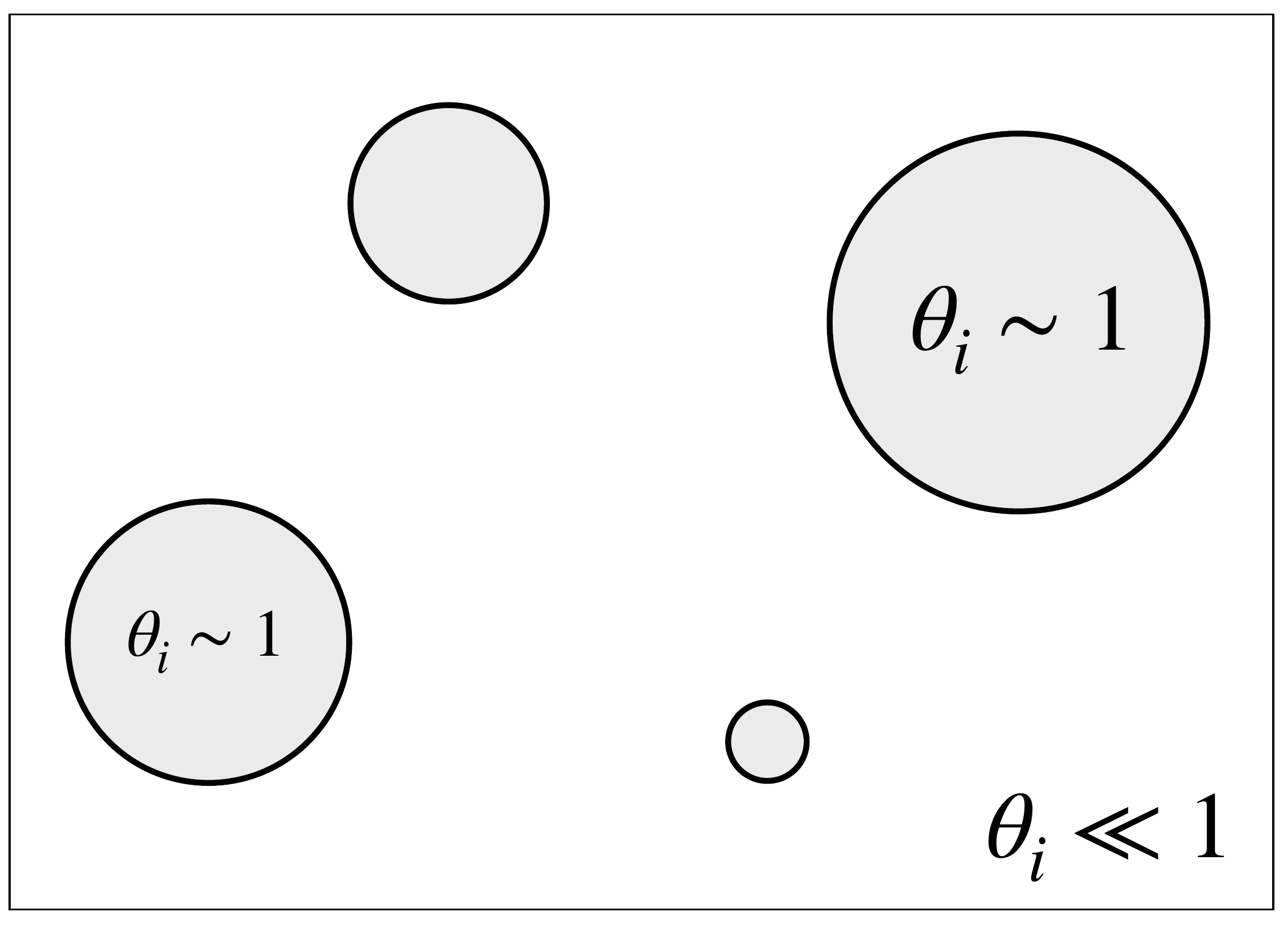}
\caption{
Schematic illustration of the axion bubbles.
}
\label{fig:ab}
\end{figure}

During inflation the axion field acquires quantum fluctuations whose distribution can be well approximated to be a Gaussian centered at about a certain field value.
Then, after the Witten effect turns on, the axion settles down at the nearest minimum. For our purpose of linking the QCD axion to PBH, the axion density must be extremely high in proto-PBH regions, and therefore,
we focus on large values of $f_a = 10^{16}$--$10^{17}$\,GeV. 
Then, one of the minima of $V_W(\phi)$ should be sufficiently close to the minimum of $V_{\rm QCD}(\phi)$ in order to get the right dark matter abundance. For later convenience, we denote $\phi_{\rm min}^{(0)}/f_a = -\epsilon$ with $0 < \epsilon \ll 1$. In spatial patches where the initial value of the axion determined by the inflationary fluctuation is smaller than the critical value $\phi_c$ (see Fig.~\ref{fig:pot}), the axion rolls down to $\phi_{\rm min}^{(0)}$.
On the other hand, in spatial patches where the initial value is larger than $\phi_c$, 
the axion rolls down to another minimum of $V_W(\phi)$, $\phi_{\rm min}^{(1)}/f_a =\pi -\epsilon$.\footnote{
The initial angle inside the bubble is close to the potential maximum because of $N_H=2$ and $\epsilon \ll 1$. 
Thus, the local axion density inside the bubble is enhanced thanks to the anharmonic effect.
Note that our scenario should work as long as the initial angle is ${\cal O}(1)$ inside the bubble and the axion can dominate over radiation
soon after it starts to oscillate. } The local axion density in this patch becomes much higher than the one with $\phi_{\rm min}^{(0)}$ after the QCD phase transition.\footnote{Precisely speaking,  the axion potential significantly deviates from a single cosine function
in the low temperature based on the standard chiral perturbation theory (see e.g. Ref.~\cite{diCortona:2015ldu}),
and we need take account of it to estimate the axion density inside the axion bubbles.  However, we have numerically checked that such modification of the potential only changes the resultant density by at most  30\% or so for the parameters of our interest. In our numerical analysis, we therefore use the simple cosine function for simplicity.}
We call such isolated rare region {\it axion bubble} in which the initial misalignment angle is set to be much larger than the background value.
See Fig.~\ref{fig:ab}.
Since the axion density in those regions soon becomes larger than the radiation density, its volume fraction should be sufficiently small for successful big bang cosmology. 
The axion bubble is
analogous to the high baryon bubble in inhomogeneous Affleck-Dine baryogenesis \cite{Dolgov:2008wu,Hasegawa:2017jtk,Hasegawa:2018yuy}.

The axion bubbles are formed when the transient PQ breaking turns on and the axion field is dynamically set at the nearest minimum, 
well before the QCD phase transition. The subsequent evolution of the axion bubbles depends on its typical size or the corresponding wavenumber. As
we shall see later, small bubbles enter the horizon well before the QCD phase transition, i.e. before the axion gets the mass from the QCD instanton effect, 
and they are dissipated away in the form of the gradient energy.\footnote{
Here we assume that the potential from the Witten effect has already disappeared. Otherwise, domain walls exist between two spatial regions in which the axion falls into one of the minima and another (e.g. $\phi_{\rm min}^{(0)}$ and $\phi_{\rm min}^{(1)}$). In this case, the bubble is surrounded by a closed domain wall and when the bubble reenters the horizon, it collapses by the tension of the domain wall and disappear in a few Hubble time by emitting nonzero mode axions.
}
On the other hand, larger axion bubbles which enter the horizon around the QCD epoch can 
collapse into PBHs or cluster into small axion clumps.

%%%%%%%%%%%%%%% FIGURE  %%%%%%%%%%%%%%%
%%%%%%%%%%%%%%%%%%%%%%%%%%%%%%%%%%%

%%%%%%%%%%%%%%% FIGURE  %%%%%%%%%%%%%%%
%%%%%%%%%%%%%%%%%%%%%%%%%%%%%%%%%%%

\subsection{PBH formation}

The formation of PBH occurs when ${\cal O}(1)$ density fluctuations, initially generated on super-horizon scales, reenter the horizon. Assuming the radiation dominated universe, the PBH mass is roughly given by the horizon mass at the formation time,
\begin{equation}
\label{eq:PBHmass}
M_{\rm PBH} = \frac{4\pi}{3} \gamma H^{-3}_f \rho_{r,f} \simeq 0.049 M_\odot \gamma \left(\frac{g_{*f}}{100} \right)^{-1/2} \left(\frac{T_f}{1\,{\rm GeV}} \right)^{-2},
\end{equation}
where $\gamma$ is ${\cal O}(1)$ numerical factor and the subscript $f$ indicates that the quantity is evaluated at the PBH formation time. 
For simplicity, we set $\gamma=1$ in what follows.

Inside axion bubbles, the energy density of the axion dominates over the radiation density much earlier than the matter-radiation equality.  Let us denote the background cosmic temperature at which the axion becomes dominant over the radiation inside bubbles by $T_B$. 
The axion mass is constant when the axion dominates inside the bubble (i.e. $T_B < T_*$) and one obtains
\begin{equation}
T_B \simeq 
3.0\,{\rm MeV} \left(\frac{f_a}{10^{16}\,{\rm GeV}} \right)^{1.2}  \left(\frac{g_{*1}}{60}\right)^{-0.42} \left[1+0.084\ln\left(\frac{f_a}{10^{16}\,{\rm GeV}}\right) \right]^{1.2},
\end{equation}
where we have neglected ${\mathcal O}(\epsilon)$ contributions and $\ln(g_{*1})$ term in the square bracket. Note that the above equation is valid for $f_a \lesssim 10^{17}$ GeV where the axion commences oscillation when the axion mass is temperature-dependent.
When the axion dominates over the radiation inside the bubble (i.e. $T<T_B$) and the size of the bubble is larger than the horizon size, 
the criterion for the PBH formation is satisfied when the bubble enters the horizon since the local energy density inside the bubble becomes significantly larger than the background radiation density.
Thus, the temperature at the axion domination inside the bubble determines the minimum mass of PBH, $M_{\rm PBH}^{\rm (min)}$, obtained by substituting $T_f = T_B$ in (\ref{eq:PBHmass}),
\begin{equation}
\label{eq:MPBHmin}
M_{\rm PBH}^{\rm (min)} \simeq 1.7\times 10^4 M_\odot \left(\frac{10^{16}\,{\rm GeV}}{f_a}\right)^{2.3} \left(\frac{g_{*1}}{60}\right)^{0.84} \left(\frac{g_{*f}}{10}\right)^{-0.5} \left[1+0.084\ln\left(\frac{f_a}{10^{16}\,{\rm GeV}}\right) \right]^{-2.3}.
\end{equation}

As the background temperature becomes lower than $T_B$, the ratio of the local energy density inside the bubble, $\rho_B^{\rm (loc)}$, to the background radiation density increases by a factor $\rho_B^{\rm (loc)}/\rho_r > 1$, 
and hence one might expect that the PBH mass can be much larger than the horizon mass defined by the background radiation.
However, as pointed out in \cite{Kopp:2010sh,Carr:2014pga}, the mass of the PBH from the collapse of those bubbles cannot be much larger than the background horizon mass. It is because the initially superhorizon-scale region in which the local energy density is significantly larger than the background value gets pinched off and separated into a baby universe through the formation of a wormhole shortly after the horizon entry, while a black hole is left behind in our universe \cite{Kopp:2010sh}. Hence, we assume that the PBH mass is given by the horizon mass of the background radiation even in the case where the PBH formation occurs after the axion domination inside the bubble, i.e. $T_f < T_B$.

\subsection{PBH abundance}
In order for our scenario to work, inflationary fluctuations necessarily distribute axion bubbles as rare objects in our universe.
It depends on the Hubble parameter during inflation, 
and the initial axion field value (the mean value in the observable patch of the Universe).
The evolution of the probability density function is governed by the Fokker-Planck equation
\begin{equation}
\frac{\partial P(N,\phi)}{\partial N} = \frac{\partial}{\partial \phi} \left[ \frac{V_\phi P(N,\phi)}{3H_{\rm inf}^2}+\frac{H_{\rm inf}^2}{8\pi^2} \frac{\partial P(N,\phi)}{\partial\phi} \right],
\end{equation}
where $N$ is the e-folding number and $H_{\rm inf}$ is the Hubble parameter during inflation.
Since the classical dynamics of the axion is negligible during inflation, the evolution is determined only by the diffusion term i.e. the second term in the square bracket in the Fokker-Planck equation. 
If the inflation was long enough before the CMB-scale fluctuations exited the horizon, we expect that all values of the initial angle between $-\pi$ and $\pi$ are equally likely. 
Since what we can observe is the dynamics after the current horizon scale exited the horizon during inflation, if we average the distribution of the angles over the scale corresponding to the current horizon
and neglect scales larger than that, we can approximate the distribution at that point $(N=0)$ as
the Dirac $\delta$-function, $P(0,\phi) = \delta(\phi-\phi_i)$.
Then, 
the solution is simply given by the Gaussian distribution function,
\begin{equation}
\label{eq:gaussianPDF}
P(N,\phi) = \frac{1}{\sqrt{2\pi}\sigma(N)} \exp\left(-\frac{(\phi-\phi_i)^2}{2\sigma^2(N)} \right) ~~\text{with}~~\sigma(N) = \frac{H_{\rm inf}}{2\pi} \sqrt{N}.
\end{equation}

\begin{figure}[tp]
\centering
\includegraphics [width=9cm,clip]{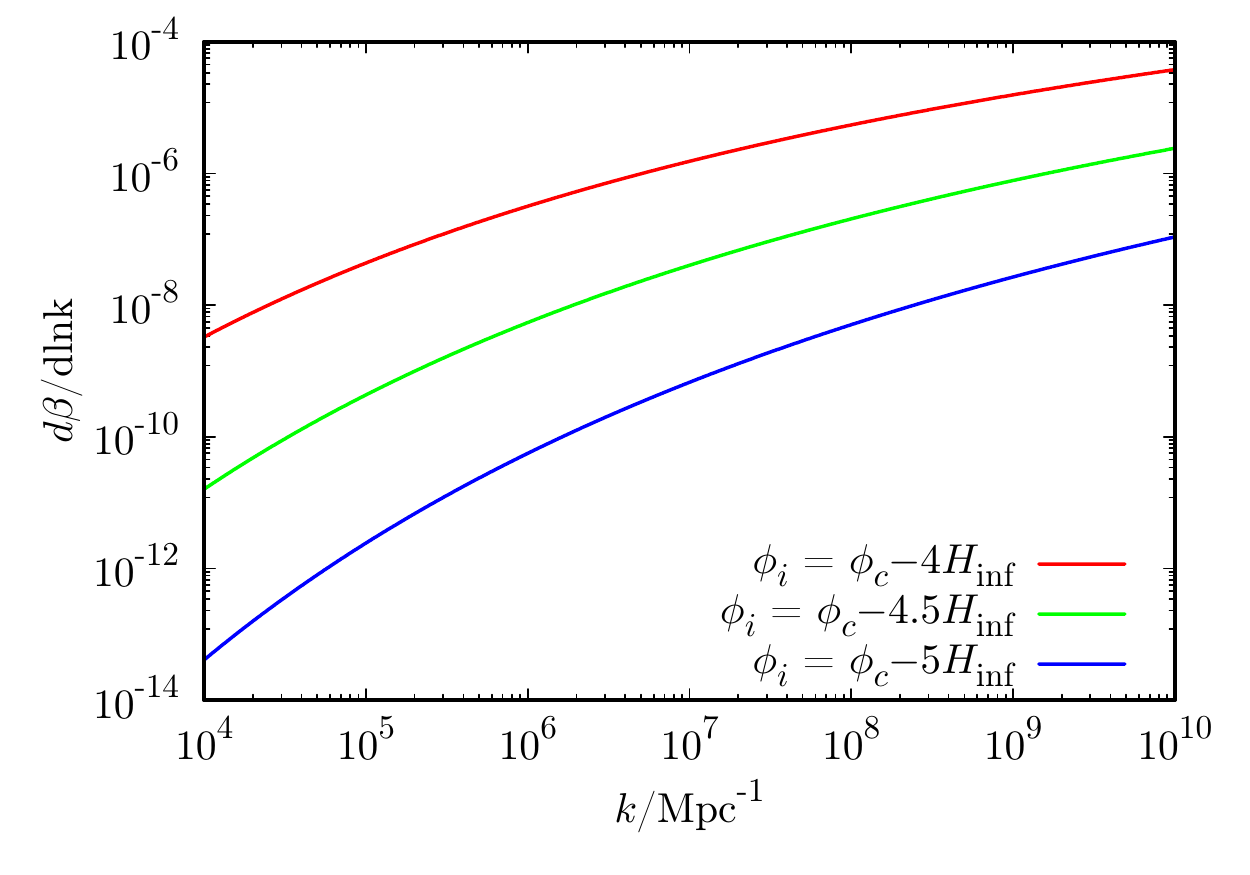}
\caption{
The volume fraction of axion bubbles with differential logarithmic interval of $k$ for $\phi_i -\phi_c=4H_{\rm inf}$ (red), $4.5H_{\rm inf}$ (green), $5H_{\rm inf}$ (blue). We have taken $N_*=0$ and $k_*=0.002\,{\rm Mpc}^{-1}$.
}
\label{fig:beta}
\end{figure}

The differential volume fraction of the axion bubble which is originally nucleated in the interval between $N$ and $N+dN$ during inflation is given by \cite{Hasegawa:2018yuy}
\begin{equation} \label{eq:dbetadN}
\frac{d\beta}{dN} = \int_{\phi_c}^{\phi_{\rm max}} \frac{\partial P(N,\phi)}{\partial N} d\phi,
\end{equation}
where $\phi_{\rm max}$ denotes the maximum value for our scenario to work.
In practice, the integrand in Eq.~(\ref{eq:dbetadN}) contributes only in the vicinity of the lower limit. Then, one can safely replace the upper limit of the integral with the infinity and the integral can be expressed by the complementary error function, namely
\begin{equation}
\frac{d\beta}{dN} \approx \frac{\partial}{\partial N}\int_{\phi_c}^{\infty} P(N,\phi)  d\phi = \frac{1}{2}\frac{\partial}{\partial N} {\rm erfc}\left(\frac{\phi_c-\phi_i}{\sqrt{2}\sigma(N)}\right) = \frac{\phi_c-\phi_i}{2} P(N,\phi_c).
\end{equation}
This reads the volume fraction of the bubbles corresponding to the scale which exits the horizon at the e-folding number $N$ during inflation.
Using $dN=d\ln k$, one obtains the differential volume fraction of the axion bubble with a logarithmic interval of the wave number as follows,
\begin{equation} \label{eq:volume_fraction}
\frac{d\beta}{d\ln k} = \frac{\phi_c-\phi_i}{2} P(\ln(k/k_*)+N_*,\phi_c),
\end{equation}
where $N_*$ and $k_*$ are some reference e-folding and wave number respectively.
Fig.~\ref{fig:beta} shows the differential volume fraction with $N_*=0$ and $k_*=0.002\,{\rm Mpc}^{-1}$.

%%%%%%%%%%%%%%% FIGURE  %%%%%%%%%%%%%%%
%%%%%%%%%%%%%%%%%%%%%%%%%%%%%%%%%%%

\begin{figure}[tp]
\centering
\includegraphics [width=9cm,clip]{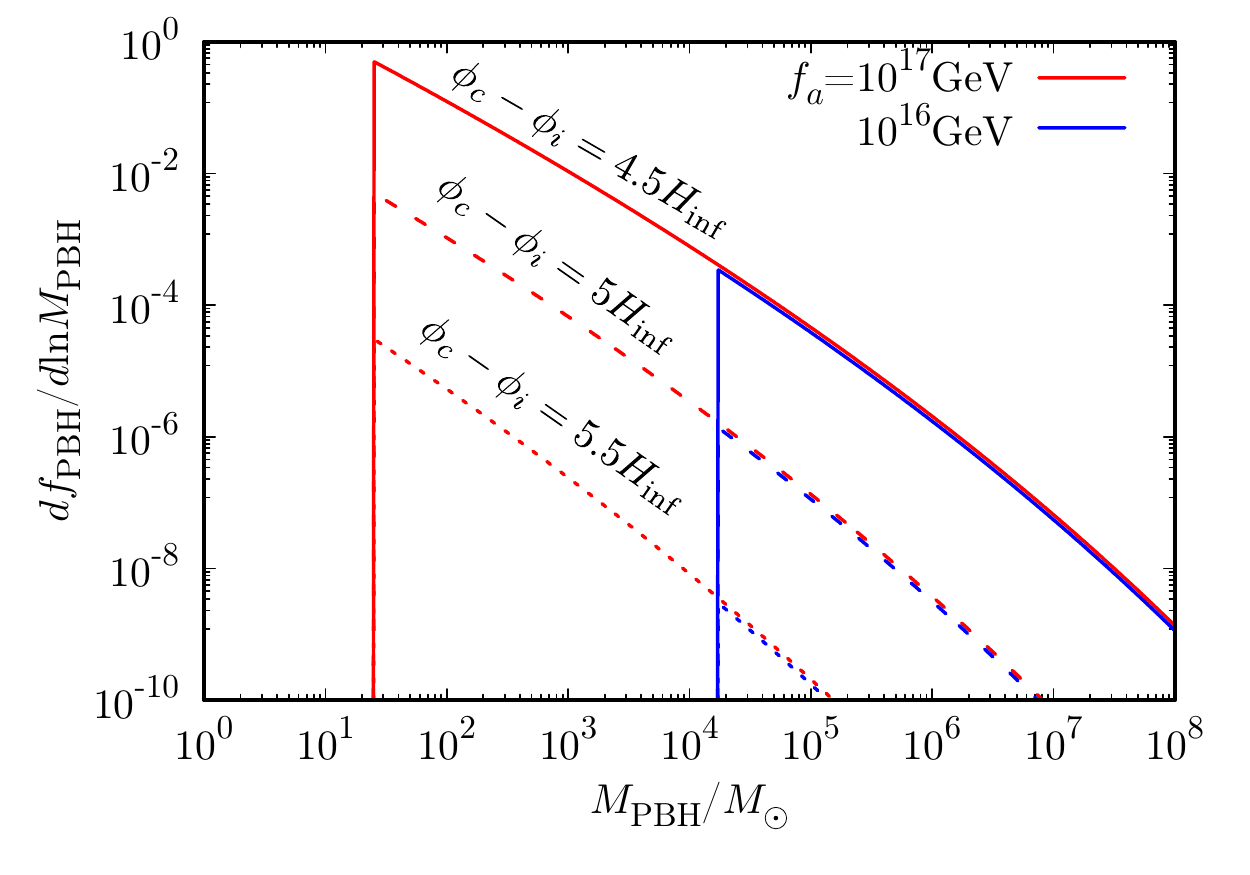}
\caption{
The predicted spectrum of the PBH fraction $f_{\rm PBH} = \Omega_{\rm PBH}/\Omega_{\rm CDM}$ with respect to the PBH mass. The solid, dashed and dotted lines correspond to $\phi_c-\phi_i=4.5H_{\rm inf}$, $5H_{\rm inf}$ and $5.5H_{\rm inf}$ respectively. We have taken $f_a = 10^{17}$ GeV (red) and $10^{16}$ GeV (blue), $N_*=0$ and $k_*=0.002\,{\rm Mpc}^{-1}$.
}
\label{fig:fPBH}
\end{figure}

Thus, one obtains the energy density of the PBH formed at $T=T_f \,(<T_B)$ divided by the entropy density,
\begin{equation}
\frac{\rho_{\rm PBH}}{s} = \frac{3}{4}T_f  \frac{d\beta}{d\ln k}\bigg|_{k=a_fH_f}.
\end{equation}
Using the relation between the PBH mass and the wavenumber corresponding to the horizon scale at the formation,
\begin{equation}
k_f \simeq 4.3\times 10^6\,{\rm Mpc}^{-1} \left(\frac{g_{*f}}{10}\right)^{-1/12} \left(\frac{M_\odot}{M_{\rm PBH}}\right)^{1/2},
\end{equation}
one can get the mass spectrum of the PBH fraction $f_{\rm PBH} = \Omega_{\rm PBH}/\Omega_{\rm CDM}$, as shown in Fig.~\ref{fig:fPBH}. Note that PBHs with the minimum mass given by (\ref{eq:MPBHmin}) are most abundantly formed because the probability to exceed the critical field value becomes larger as the corresponding scale becomes smaller (see (\ref{eq:gaussianPDF})).
 Hence, the typical mass of the PBH in this scenario is given by the minimum mass (\ref{eq:MPBHmin}), which has one-to-one correspondence to the decay constant $f_a$ if the axion accounts for the dominant component of the dark matter. 

Note also that the typical mass of the PBH is ${\cal O}(10)M_\odot$ for $f_a = 10^{17}$GeV. However, the CMB bounds due to the gas accretion onto PBHs put a stringent upper bound on the PBH fraction within the mass range $10 M_\odot \lesssim M_{\rm PBH} \lesssim 10^4 M_\odot$; specifically, it is $f_{\rm PBH} \lesssim 3 \times 10^{-9}$ at $M_{\rm PBH} \sim 10^4 M_\odot$ \cite{Serpico:2020ehh}. Taking into account this constraint, the PBH fraction with mass ${\cal O}(10)M_\odot$ should satisfy $f_{\rm PBH} \lesssim 10^{-5}$ in our scenario.
It is smaller than the required value in Ref.~\cite{Sasaki:2016jop} to explain the LIGO/Virgo events. However, our scenario potentially predicts initial clustering of PBHs because large-scale fluctuations (of the axion field value as an isocurvature mode) are accumulated to make the fluctuation large enough for the PBH formation and thus there exist sizable mode-mode couplings. In such a case, the merger rate of binary PBHs can be significantly enhanced and the LIGO/Virgo events can be explained even if the PBH fraction is as small as $f_{\rm PBH} \sim 10^{-7}$ \cite{Young:2019gfc}. To deduce whether our scenario can explain the observed merger events, the clustering property should be investigated quantitatively and it is left for future work.  If the binary formation rate can be enhanced by more than a factor of $100$,
 our scenario for $f_a \sim 10^{17}$ GeV can explain simultaneously the PBH as the seed of SMBH with mass ${\cal O}(10^4)M_\odot$ and the responsible for observed binary merger events with mass ${\cal O}(10)M_\odot$.

%%%%%%%%%%%%%%% FIGURE  %%%%%%%%%%%%%%%
%%%%%%%%%%%%%%%%%%%%%%%%%%%%%%%%%%%

\subsection{Bubble abundance}

In the previous two sections, we have focused only on the PBH formation. In fact, our scenario also predicts smaller bubbles which enter the horizon at $T>T_B$ before the axion becomes dominant inside them. The fate of such small bubbles is divided into the following two cases. 
First, if the local axion density inside the bubble is larger than the background value, the bubble eventually evolves into the axion minicluster. Second, if the size of the bubble enters the horizon well before the QCD epoch, axions inside bubbles are dissipated away in the form of relativistic particles.

Here, let us consider the relative density contrast of the axion inside the bubble to the background one. If the bubble size is larger than the Hubble horizon at the commencement of the background axion oscillation (at $T=T_1$), the ratio of the local axion density inside the bubble, denoted by $\rho_B^{\rm (loc)}$, to the background density $\rho_a$ is 
\begin{equation} \label{eq:rhoBloc-rhoa}
\frac{\rho_B^{\rm (loc)}}{\rho_a} = \frac{(n_a/s)_{\theta_i = \pi-\epsilon}}{(n_a/s)_{\theta_i = -\epsilon}} \simeq 5.2 \times 10^6\left(\frac{f_a}{10^{16}\,{\rm GeV}}\right)^{1.2} \left(\frac{g_{*1}}{60} \right)^{-0.42} \left[1+0.084\ln\left(\frac{f_a}{10^{16}\,{\rm GeV}}\right) \right]^{1.2},
\end{equation}
where we have substituted the observed CDM abundance for the background axion density.
Note that the global energy density of the axion in the bubble is given by $\rho_B = \rho_B^{\rm (loc)} \beta$ with $\beta$ the total volume fraction of the axion bubbles.
Hence, one obtains the spectrum of the energy density fraction of the axion bubble by multiplying the above ratio by the differential volume fraction (\ref{eq:volume_fraction}),
\begin{equation}
\frac{1}{\rho_a}\frac{d\rho_B}{d\ln k} = \frac{\rho_B^{\rm (loc)}}{\rho_a} \frac{d\beta}{d\ln k}.
\end{equation}
Note that the above estimation is valid for wavenumber smaller than $k_1 = a(T_1)H(T_1)$ given by
\begin{equation} \label{eq:k1}
k_1 \simeq 3.1 \times 10^6 \,{\rm Mpc}^{-1}\left(\frac{g_{*1}}{60}\right)^{0.084} \left(\frac{10^{16}{\rm GeV}}{f_a}\right)^{0.16}.
\end{equation}

Those bubbles which enter the horizon 
before the axion inside them dominates over the radiation (i.e. $T > T_B$)
cannot collapse into PBHs. Instead, they form small clumps called axion miniclusters \cite{Hogan:1988mp}. The cluster mass is roughly given by the energy density of the axion inside the bubble multiplied by the horizon volume. For bubbles whose size enters the horizon after the onset of the axion oscillation (i.e. $k<k_1$), one can estimate the minicluster mass as follows,
\begin{equation} \label{eq:Mmc}
M_{\rm mc} \simeq 0.020 M_\odot \left(\frac{f_a}{10^{16}\,{\rm GeV}}\right)^{1.7} \left(\frac{k_1}{k}\right)^3 \left(\frac{g_{*1}}{60} \right)^{-0.67}\left[ 1+0.084 \ln\left(\frac{f_a}{10^{16}\,{\rm GeV}}\right) \right]^{1.2}.
\end{equation}

\begin{figure}[tp]
\centering
\subfigure[]{
\includegraphics [width = 7.5cm, clip]{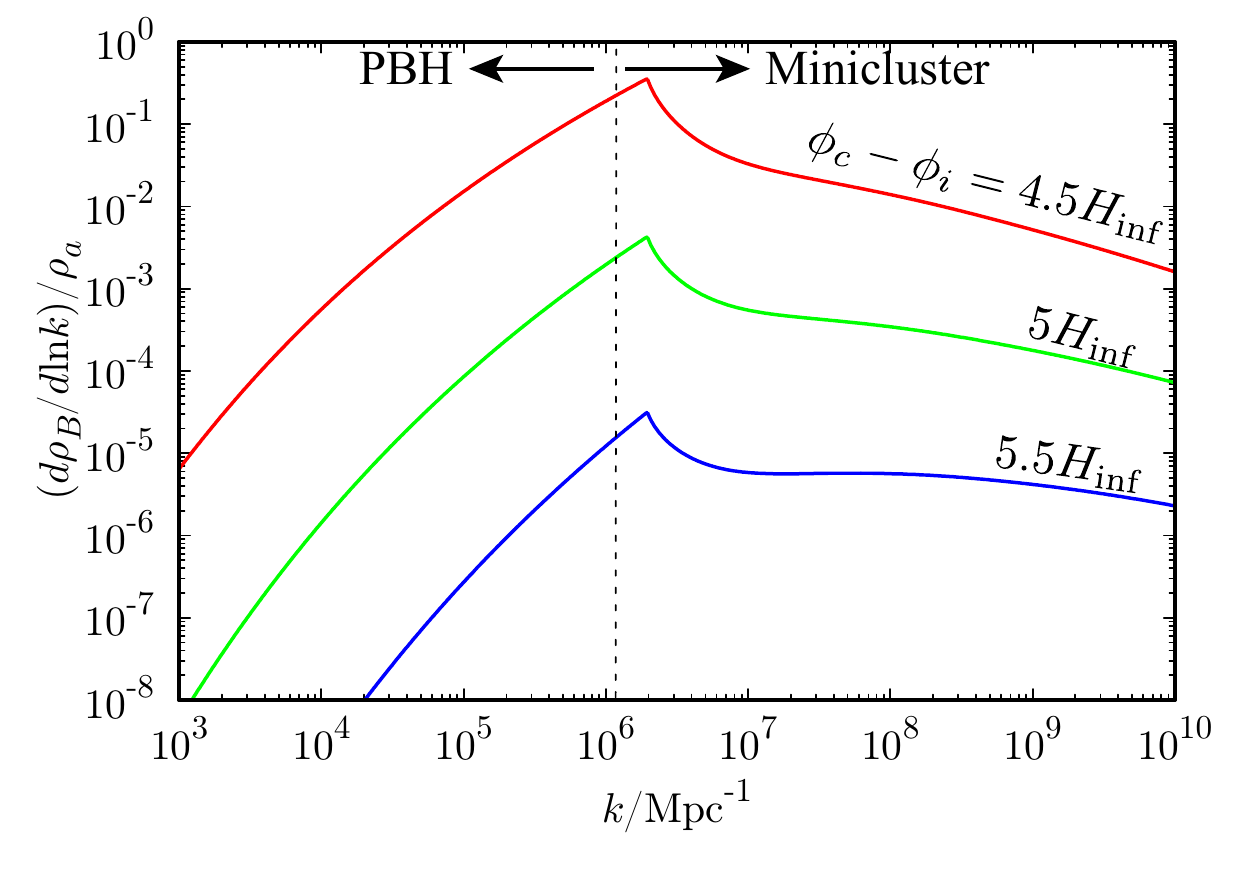}
\label{subfig:bubble_fraction_17}
}
\subfigure[]{
\includegraphics [width = 7.5cm, clip]{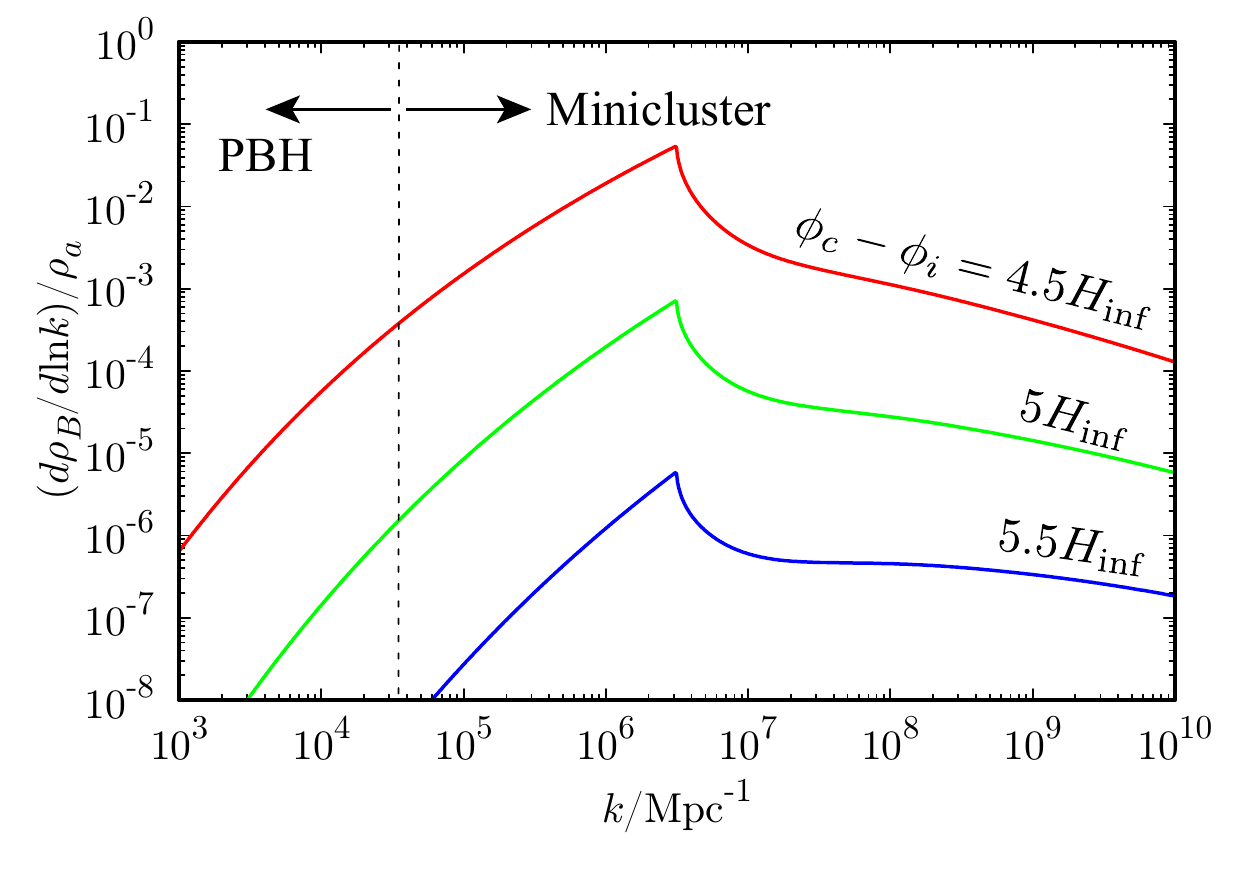}
\label{subfig:bubble_fraction_16}
}
\caption{
The spectrum of the energy density fraction of the axion bubble to the background axion density. We have taken $f_a = 10^{17}$ GeV (left), $10^{16}$ GeV (right) and $\phi_c-\phi_i = 4.5H_{\rm inf}$ (red), $5H_{\rm inf}$ (green) and $5.5H_{\rm inf}$ (blue).
}
\label{fig:bubble_fraction}
\end{figure}

Now let us consider smaller bubbles whose physical size is smaller than $H(T_1)^{-1}$ at $T_1$.
Here, we neglect the transient potential. 
In this case, when the bubble size enters the horizon, the axion is massless but the bubble stores the gradient energy $\rho_{\rm grad} \sim (H \pi f_a)^2/2$. Since the gradient energy decays in the same way as the relativistic component, the field value of such nonzero-mode axion decays inversely proportional to the scale factor.
Thus, the ratio (\ref{eq:rhoBloc-rhoa}) is determined by the field amplitude of such nonzero-mode axions at $T=T_1$ as a function of the wavenumber,
\begin{equation}
\theta_1 \sim \pi \left(\frac{m_a(T_1)}{3H_k} \right)^{1/2}  \simeq 1.4 g_{*k}^{-1/12} \left(\frac{g_{*1}}{60}\right)^{0.17} \left(\frac{f_a}{10^{16}\,{\rm GeV}} \right)^{-0.16}\left(\frac{10^7\,{\rm Mpc}^{-1}}{k} \right),
\end{equation}
where $H_k$ and $g_{*k}$ are respectively the Hubble parameter and the relativistic degrees of freedom when the mode $k$ reenters the horizon. Hence, one obtains the resultant density fraction of the axion from those small bubbles by replacing $\theta_i =  \pi-\epsilon \to \theta_1$ and $m_a(T_1) \to \sqrt{(k/a(T_1))^2+m_a^2(T_1)}$ for $n_a/s$ in the numerator in (\ref{eq:volume_fraction}).
Note that such nonzero-mode axions also form minicluster as long as their local density is larger than the background value. Thus, by equating local density of the nonzero-mode axion and the density of the background axion at $T=T_1$, one can obtain the minimum minicluster mass,
\begin{equation}
M_{\rm mc,min} \simeq 3.0 \times 10^{-9} M_\odot \left( \frac{f_a}{10^{16}\,{\rm GeV}} \right)^{1.7} \left( \frac{g_{*1}}{60} \right)^{-0.25},
\end{equation}
and corresponding maximum wave number is 
\begin{equation}
k_{\rm mc,min} \simeq 0.95 \times 10^{12}\, {\rm Mpc}^{-1} g_{*k}^{-1/8} \left(\frac{f_a}{10^{16}\,{\rm GeV}}\right)^{-0.16} \left(\frac{g_{*1}}{60} \right)^{-0.21}.
\end{equation}
The resultant spectrum of the bubble fraction is shown in Fig.~\ref{fig:bubble_fraction}. In the figure, the peak wavenumber is $k_1$ given by (\ref{eq:k1}) corresponding to the scale reentering the horizon at the onset of the axion oscillation and thus the minicluster formed at that time is the most abundant. This is because the amplitude of the axion oscillation has not been damped and the anharmonic effect efficiently enhance the local axion density inside those bubbles. Therefore, the mass of the most abundant minicluster is given by (\ref{eq:Mmc}) with $k=k_1$.

%%%%%%%%%%%%%%% MULTI-FIGURE  %%%%%%%%%%%%%%%
%%%%%%%%%%%%%%%%%%%%%%%%%%%%%%%%%%%%%%%

%%%%%%%%%%%%%%%%%%%%%%%%%%%%%%%%%%%%%%%%%%%%%%%%%%%%%%%%%%%%%%%%%%%%%%
\section{Conclusions and discussions} \label{sec:conc}
%%%%%%%%%%%%%%%%%%%%%%%%%%%%%%%%%%%%%%%%%%%%%%%%%%%%%%%%%%%%%%%%%%%%%%

We have shown that a significant number of PBHs with mass $M_{\rm PBH} \gtrsim 10\,M_\odot$ can be formed from the QCD axion with $f_a \sim 10^{17}$ GeV whose early dynamics is modified by the temporally large PQ breaking term. As a specific example, we have considered the Witten effect which 
generates an additional contribution to the axion potential. Then, the axion starts to oscillate around the nearest minimum even before the QCD phase transition. In general, the potential from the Witten effect has multiple minima. This implies that the subsequent evolution of the universe can be rather different 
depending on which minimum the axion settles down.
We have found that the axion dynamics leads to copious formation of the high density axion bubbles, which collapse into PBHs or lead to miniclusters 
after the QCD phase transition. Interestingly, the mass of the PBHs is determined by the QCD scale, which is roughly $M_{\rm PBH} \gtrsim 10 (10^4)M_\odot$ for 
$f_a \sim 10^{17}(10^{16})$ GeV. Thus, one can explain the LIGO event and/or the seeds of SMBH by those PBHs, while the QCD axion with large $f_a$
constitute the observed dark matter.

The PBH in this mass range can be testable by future 21cm observation by e.g. the SKA telescope \cite{Gong:2017sie,Gong:2018sos,Mena:2019nhm}.
Note that the constraint due to induced gravitational waves by the pulsar timing observation can be evaded \cite{Saito:2008jc}.
In addition, the QCD axion with relatively large decay constant ($f_a \gtrsim 10^{16}$--$10^{17}$ GeV) can be searched for by direct detection experiments such as ABRACADABRA \cite{Kahn:2016aff}. The axion minicluster is also a potential target to test our scenario. 
In particular, our scenario predicts heavier minicluster than that in the conventional scenario where the post-inflationary PQ symmetry breaking is followed by the formation and decay of topological defects \cite{Fairbairn:2017dmf} and thus motivates upgrade of microlensing survey beyond the current upper bound put by EROS \cite{Tisserand:2006zx}, HSC \cite{Niikura:2017zjd} and OGLE \cite{Niikura:2019kqi}.

%%%%%%%%%%%%%%%%%%%%%%%%%%%%%%%%%%%%
\section*{Acknowledgment}
We thank Masaki Yamada for helpful comments.
This work is supported by JSPS KAKENHI Grant Numbers
JP15H05889 (F.T.), JP15K21733 (F.T.),  JP17H02875 (F.T.),
JP17H02878 (F.T.), JP18H01243 (N.K.), JP19K14708 (N.K.), JP19H01894 (N.K.), 20H01894 (N.K. and F.T.) and by World Premier International Research Center Initiative (WPI Initiative), MEXT, Japan.
%%%%%%%%%%%%%%%%%%%%%%%%%%%%%%%%%%%%

\bibliographystyle{utphys}
\bibliography{ref}

\end{document}